\documentclass{IEEEtran}

\usepackage{xcolor}
\ifCLASSINFOpdf
   \usepackage[pdftex]{graphicx}
\else
   \usepackage[dvips]{graphicx}
\fi
\usepackage[cmex10]{amsmath}

\usepackage[ruled,vlined]{algorithm2e}
\usepackage{amssymb}  
\usepackage{bm}  
\usepackage{hyperref}

\newcommand{\Y}{\mathcal{Y}}  
\newcommand{\E}[1]{\mathrm{E} \left[ #1 \right]}  

\DeclareMathOperator*{\argmin}{argmin}

\begin{document}
%
\title{A Bayesian Framework for Power System Components Identification}


\author{\IEEEauthorblockN{Artem Mikhalev\IEEEauthorrefmark{1},
Alexander Emchinov\IEEEauthorrefmark{1},
Samuel Chevalier\IEEEauthorrefmark{2}, 
Yury Maximov\IEEEauthorrefmark{3}\IEEEauthorrefmark{4} and
Petr Vorobev\IEEEauthorrefmark{4}}\\
\IEEEauthorblockA{\IEEEauthorrefmark{1}  School of Applied Mathematics and Informatics, 
Moscow Institute of Physics and Technology}\\  
\IEEEauthorblockA{\IEEEauthorrefmark{2} Department of Mechanical Engineering, 
Massachusetts Institute of Technology}\\  
\IEEEauthorblockA{\IEEEauthorrefmark{3} Center for Nonlinear Studies and Theoretical Division, 
Los Alamos National Laboratory}\\
\IEEEauthorblockA{\IEEEauthorrefmark{4} Center for Energy Science and Technology, 
Skolkovo Institute of Science and Technology} 
}

\maketitle



\begin{abstract}
Having actual models for power system components (such as generators and loads or auxiliary equipment) is vital to correctly assess the power system operating state and to establish stability margins. However, power system operators often have limited information about the actual values for power system component parameters. Even when a model is available, its operating parameters and control settings are time-dependent and subject to real-time identification. Ideally, these parameters should be identified from measurement data, such as phasor measurement unit (PMU) signals. However, it is challenging to do this from the ambient measurements in the absence of transient dynamics since the signal-to-noise ratio (SNR) for such signals is not necessarily large.  In this paper, we design a Bayesian framework for on-line identification of power system component parameters based on ambient PMU data, which has reliable performance for SNR as low as five and for certain parameters can give good estimations even for unit SNR. We support the framework with a robust and time-efficient numerical method. We illustrate the approach efficiency on a synchronous generator example. 
\end{abstract}





\section{Introduction}
Accurate and dynamically updated model information for power system components can help a system operator improve power system performance in multiple ways---accurate stability margin assessment, better reserve allocation, and identification of perturbations (such as low-frequency oscillations) are just a few examples. Although there exists a body of literature dedicated to model estimation methods from phasor measurement unit (PMU) data following different transients, methods for model inference from ambient data are less developed. In this manuscript, we present a Bayesian framework method for estimating model parameters from ambient PMU data that can potentially be applied to various power grid components.

Performing parameter inference from observed PMU data is generally considered to be a so-called inverse problem~\cite{Hiskens:2001,Hiskens:1999}. Due to the nonlinear nature of power systems and the unsolved problem of load modeling, solving inverse problems in power systems is a challenging task. Much of the literature focuses on parameter estimation following some large transient disturbance. In papers~\cite{Zhu:2018,Choi:2006}, for example, the authors perform load parameter estimation based on a composite load model after a switching event, but a model selection is incorporated with the parameter selection procedure. The authors of~\cite{Guo:2014} use modal identification to estimate system eigenvalues and use sensitivity analysis to infer system parameters, but the methods are primarily designed to infer aggregate generator inertias. Other proposed inference methods incorporate a prior analytical model based on the physics of the underlying system components, such as~\cite{Petra:2017,bayesian_framework,Haifeng:2019,gorbunov2019estimation}, but the results of~\cite{bayesian_framework} have been developed explicitly for forced oscillation identification. Papers~\cite{Petra:2017} and~\cite{Haifeng:2019} cover arbitrary nonlinear dynamics in the time domain and thus are fairly computationally intensive.

Because most of the methods are focused on using measurements from dynamic transients to estimate power system parameters, it is assumed the signal-to-noise ratio (SNR) is fairly large. This, however, is not necessarily the case for ambient fluctuations of the system states. Thus, for the goal of estimating power system parameters from ambient PMU data, one has to develop a method that performs well even under considerably small SNR values. In this manuscript, we present a method for identification of power system component parameters from the measurements of ambient fluctuations of bus voltage and current. We extend our earlier-introduced Bayesian framework \cite{bayesian_framework} to perform under reasonably small SNR values by employing state-of-the-art cross-entropy optimization methods when performing the maximum likelihood procedure. We illustrate the performance of our method on a generator model, for which we estimate its parameters, as they are not accurately known.    
\section{Problem Formulation}\label{sec:setup}
We consider a generator with PMU measurements of voltage and current available on its terminals. Our goal is to design a method that identifies the unknown parameters of the generator model from the ambient data of PMU measurements. This can be done by applying Bayes rule to the measurement data, since voltage and current fluctuations are related to each other through the generator model. Then, comparing the measured values of current to the expected values, one can infer the true parameters of the underlying model. However, there is always measurement noise present, so one can only infer the probability density functions for parameter values, which become more pronounced with a decrease in measurement noise. In this paper, we exploit the Bayes rule to infer the posterior probability density functions for generator parameters. 

Because the inference is made from ambient data, we can use linear models to describe the system dynamics; the amplitude of ambient fluctuations of voltage and current are typically rather small. Most power grid components can be described as a dynamic system in a general form:
\begin{subequations}\label{dyn_main}
    \begin{equation}
        \mathbf{\dot{x}}(t) = A(\Theta) \mathbf{x}(t) + B(\Theta) \mathbf{u}(t), 
    \end{equation}
    \begin{equation}
        \mathbf{y}(t) = C(\Theta) \mathbf{x}(t) + D(\Theta) \mathbf{u}(t),
    \end{equation}
\end{subequations}
where $\mathbf{x}$ is the vector of system states and $\mathbf{u}$ and $\mathbf{y}$ are the vectors of inputs and outputs, respectively. All represent the small-signal variations of corresponding variables around their steady-state values. $\Theta$ denotes the set of model parameters, which could be uncertain. In the case of power system components, $\mathbf{u}$ and $\mathbf{y}$ can represent the vectors of terminal voltage and current fluctuations, respectively. Both vectors are two-dimensional and can be represented either in rectangular or polar coordinates. 

The general Bayes approach includes specifying some prior probability density function $p_{\text{prior}}(\Theta)$ for the values of parameters $\Theta$ and then the a-posteriory probability density $p_{\text{post}}(\Theta |\mathbf{u}, \mathbf{y})$ using measurement data according to the following formula:
\begin{equation}\label{Bayes_rule}
    p_{\text{post}}(\Theta |\mathbf{u}, \mathbf{y})
    \propto
    p_{\text{likely}}(\mathbf{u}, \mathbf{y}| \Theta) p_{\text{prior}}(\Theta).
\end{equation}

To perform the Bayesian inference, we start by passing from a time domain to the frequency domain by performing Fourier transform on the dynamic system variables. For example, for system states $x_i$, 
\begin{equation}\label{fourier}
    x_i(\Omega) = \int_{-\infty}^{+\infty} x_i(t) e^{-j \Omega t} \mathop{dt}, \quad i \in \{ 1, \dots, n \},
\end{equation}
and likewise for inputs $u_i$ and outputs $y_i$. We note that this frequency domain has nothing to do with AC frequency; we consider we are already in a phasor domain, and dynamics given by Eq. \eqref{dyn_main} are much slower than the AC frequency. 

Because of the linear form of Eq. \eqref{dyn_main}, there exists a linear relation between the system inputs $\mathbf{u}$ and outputs $\mathbf{y}$. If we choose the vector of voltage as inputs and the vector of current as output, this relation can be written in the frequency domain using the effective generator admittance matrix:
\begin{equation}\label{y=Yu}
    \mathbf{y}(\Omega) = \Y(\Omega, \Theta) \mathbf{u}(\Omega). \quad
\end{equation}
The expression for matrix $\mathbf{y}(\Omega)$ is rather cumbersome and can be obtained from a generator model; explicit expression can be found in \cite{gen_impedance}. Here, both vectors $\mathbf{v}$ and $\mathbf{i}$ can be written in any representation, with the standard $d-q$ representation being the most wide-spread. In this paper, we will use polar representation for both voltage and current, thus Eq. \eqref{y=Yu} can be written in explicitly: 
\begin{equation}\label{IVpolar}
     \begin{bmatrix}
        I \\
        \phi \\
    \end{bmatrix} = \begin{bmatrix}
        \Y_{11} &  \Y_{12} \\
        \Y_{21} &  \Y_{22}
    \end{bmatrix} \begin{bmatrix}
        V \\
        \theta \\
    \end{bmatrix}. \quad 
\end{equation}
$I$, $\phi$, $V$, and $\theta$ are the (small-signal) current, voltage amplitude, and phase, respectively. All the components of matrix $\mathbf{Y}$ are functions of frequency and generator parameters. Their explicit expressions are rather cumbersome and can be obtained from the generator model; they can be found in \cite{gen_impedance}. Throughout the paper we will continue using vector denotations as in \eqref{y=Yu}, assuming the polar representation. 

For an ideal system, generator parameter identification can be performed by inferring the admittance matrix $\mathbf{Y}(\Omega)$ from measurements of both voltage and current. However, due to measurement noise, such a procedure cannot be performed exactly. Instead, under certain assumptions about the measurement noise properties, one can infer the probabilistic estimations for true generator parameters.  

We start by writing the measured voltage and current vectors (which we denote with a subscript $m$) as sums of true values and measurement noise: 
\begin{equation}
    \mathbf{u}_m(t) = \mathbf{u}(t) + \bm{\epsilon}(t),
    \quad
    \mathbf{y}_m(t) = \mathbf{y}(t) + \bm{\eta}(t).
\end{equation}
Here, $\mathbf{u}_m(t)$ and $\mathbf{y}_m(t)$ represent the measured voltage and current fluctuations, and $\bm{\epsilon}$ and $\bm{\eta}$ represent their corresponding measurement noise vectors, which we assume to be white. We note that since both vectors $\mathbf{u}$ and $\mathbf{y}$ represent ambient fluctuations of voltage and current around their steady-state values, the noise terms are not necessarily small comparatively. 

Next, we pass to frequency domain; however, because the measurement data is sampled as a discrete series, we will perform Discrete Fourier Transform. We assume there are $2K+1$ samples in the time domain for every dynamic variable, so that it will correspond to $K+1$ complex components in the frequency domain $\Omega_{\omega} = \{ \Omega_0, \dots, \Omega_K \}$, which we assume to be equally spaced. Thus, Eq. \eqref{y=Yu} can be written in the frequency domain as
\begin{equation}\label{y_Yu_noise}
    \mathbf{y}_m(\Omega) - \bm{\eta}
    =     \Y(\Omega, \Theta) \mathbf{u}_m(\Omega) - \Y(\Omega, \Theta) \bm{\epsilon}, \quad  \Omega \in \Omega_{\omega}.
\end{equation}
We note that since we assume the white measurement noise, both $\bm{\eta}$ and $\bm{\epsilon}$ do not depend on frequency. We also note that due to the complex-valued nature of the transformation \eqref{fourier}, components of vectors in Eq. \eqref{y_Yu_noise} become complex. Because all the vectors in the time domain were two-dimensional, each vector in Eq. \eqref{y_Yu_noise} has $2(K+1)$ complex components, or equivalently $4(K+1)$ real components.

Now, we rewrite expression \eqref{y_Yu_noise} as $\mathbf{r}=\mathbf{q}$ by moving all the measured variables to the left side and the noise variables to the right side. The vectors $\mathbf{r}$ and $\mathbf{q}$ are given by the following expressions:  
\begin{equation}\label{r_vec}
    \mathbf{r} = \mathbf{y}_m(\Omega) - \Y(\Omega, \Theta) \mathbf{u}_m(\Omega),
\end{equation}
\begin{equation}\label{q_vec}
    \mathbf{q} = \bm{\eta} - \Y(\Omega, \Theta) \bm{\epsilon}.
\end{equation}
Vector $\mathbf{r}$ now contains the measured variables, and vector $\mathbf{q}$ contains the noise variables.   

To continue building the likelihood function, vector $\mathbf{q}$ should be considered as a vector containing $4(K + 1)$ random variables from the normal distribution (with zero means). Moreover, $\mathbf{q}$ possesses multivariate normal distribution because all its components are independent from each other (as mentioned above, noise in the frequency domain doesn't depend on frequency). The $4(K + 1) \times 4(K + 1)$ covariance matrix $\Gamma_{\mathbf{q}}$ is
\begin{equation}\label{Gamma_L}
    \Gamma_{\mathbf{q}} = 
    \begin{bmatrix}
        \Gamma_{q_{r} q_{r}} & \Gamma_{q_{r} q_{i}}\\
        \Gamma_{q_{i} q_{r}} & \Gamma_{q_{i} q_{i}}
    \end{bmatrix} .
\end{equation}
Each of the submatrix in \eqref{Gamma_L} will be diagonal because measurement noise at the $k^{\text{th}}$ frequency is uncorrelated with everything except for itself at the particular $k^{\text{th}}$ frequency; that is,
\begin{equation}
\E{ \mathbf{\epsilon}_r \mathbf{\epsilon}_r^T} = \E{ \mathbf{\epsilon}_i \mathbf{\epsilon}_i^T} = \mathbf{I} 
\end{equation}
\begin{equation}
\E{ \mathbf{\epsilon}_r \mathbf{\epsilon}_i^T} = 0,
\end{equation}
and likewise for $\mathbf{\eta}$. Using these relations and Eq. \eqref{q_vec}, it is now possible to express all the components of the covariance matrix $\Gamma_{\mathbf{q}}$ in terms of the admittance matrix  $\Y(\Omega, \Theta)$. The explicit expressions are rather cumbersome, so we refer the reader to \cite{bayesian_framework}. 

The multivariate Gaussian likelihood function for $\mathbf{q}$ may now be constructed as
\begin{equation}\label{lhood}
    p_{\text{likely}} (\mathbf{y}, \mathbf{u} | \Theta) =
    \frac{ \exp \{ -\frac{1}{2} \mathbf{q}^{\! \top} \Gamma_{\mathbf{q}}^{-1} \mathbf{q} \} }{ \sqrt{ (2 \pi)^{4(K+1)} \det{\Gamma_{\mathbf{q}}} } } .
\end{equation}

We also consider the Gaussian prior distribution for the system parameters $\Theta$: 
\begin{equation}\label{pri}
    p_{\text{prior}} (\Theta) = \frac{ \exp \{ -\frac{1}{2} (\Theta - \Theta_{\text{prior}})^{\! \top} \Gamma_{\Theta}^{-1} (\Theta - \Theta_{\text{prior}}) \} }{ \sqrt{ (2 \pi)^{l} \det{\Gamma_{\Theta}} } },
\end{equation}
where $l$ is the number of uncertain system parameters.

By substituting Eqs.~\eqref{lhood} and \eqref{pri} into Eq.~\eqref{Bayes_rule}, taking a negative logarithm, assuming the determinant of the covariance
matrix $\Gamma_{\mathbf{L}}$ is roughly constant
across the plausible system parameters, and maximizing $p_{\text{post}}(\Theta |\mathbf{u}, \mathbf{y})$,
we eventually obtain an optimization problem to solve:
\begin{equation}\label{opt_problem}
\begin{split}
    \Theta_{\text{post}} = \argmin_{\Theta \in \mathbb{q}^l}
    \Big\{
        & \mathbf{q}^{\! \top} \Gamma_{\mathbf{q}}^{-1} \mathbf{q} + \\
        & + (\Theta - \Theta_{\text{prior}})^{\! \top} {\Gamma_{\Theta}^{-1}} (\Theta - \Theta_{\text{prior}})
    \Big\} .
\end{split}
\end{equation}

\section{Optimization Procedure}\label{sec:opt}

The setup of our algorithm for power system component identification in the Bayesian  framework is given in Alg.~\ref{bf_algo}.

To use the framework, one should specify input data $U$, output data $Y$, and the prior parameters $\Theta_{\text{prior}}$. (Note, a wise choice of the prior value for the system parameters decreases the computational complexity of the algorithm.) 

Arrays $Y$ (of voltage amplitudes and phases)  and $U$ (of current amplitudes and phases) have the size $2 \times (2K + 1)$, where $2K+1$ is the number of samples in the time domain for every dynamic variable.

The minimization problem given by~Eq.~\eqref{opt_problem} is non-convex in optimization parameters $\Theta$ and potentially has a lot of spurious local minima where a numerical method can be trapped. This is essential for first- and second-order optimization algorithms that use a gradient and a Hessian, respectively, to update the value of the parameter. We also note that gradient and Hessian computations are significantly more time-consuming for the function given by Eq.~\eqref{opt_problem} than the calculation of the function value. Another significant disadvantage of first- and second-order optimization schemes is their numerical instability for high SNRs due to unreliable computations of the gradient and the Hessian. 

\begin{algorithm}[!t]
\SetAlgoLined
    \KwIn{inputs $U$ (voltages and voltage phases), \par 
    \qquad outputs $Y$ (currents and current phases), \par \qquad prior value of the system parameters $\Theta_{\text{prior}}$,
    \par \qquad admittance matrix $\Y$
    }
    \KwOut{posterior value of the system parameters $\Theta_{\text{post}}$}
\nl    \qquad $\tilde{Y}$ and $\tilde{U}$ = the DFT for each row of $Y$ and $U$, \par 
\nl    \qquad Minimize the objective function, Eq.~\eqref{opt_problem}, using \par
       \qquad $\tilde{Y}$, $\tilde{U}$, $\Y$, and $\Theta_{\text{prior}}$ to obtain $\Theta_{\text{post}}$, \par
\nl    \qquad Return $\Theta_{\text{post}}$
\caption{Bayesian setup to the power systems components identification.}
\label{bf_algo}
\end{algorithm}

In this way, we suggest using the cross-entropy method, which is known to be a state-of-the-art method in zero-order optimization \cite{rubinstein1999cross, ce_optimal_rates,nemirovsky1983problem}. The method consists of a sequential generation of random points over the parameters domain, and a wise update of the parameters of sampling distribution with the aim to provide more samples in the regions with sufficiently low values of the objective function. The method is known to be especially efficient in finding an optimal value for functions with a high sensitivity of the function value to the values of its parameters \cite{mackay2019inference}.  Algorithm~\ref{ce_algo} contains details of the cross-entropy method.  Implementation details and the source code are accessible on GitHub\footnote{Implementation is available at \url{https://github.com/greylord1996/MAP}}.

\begin{figure}[!t]
\centering
\includegraphics[width=0.49\textwidth]{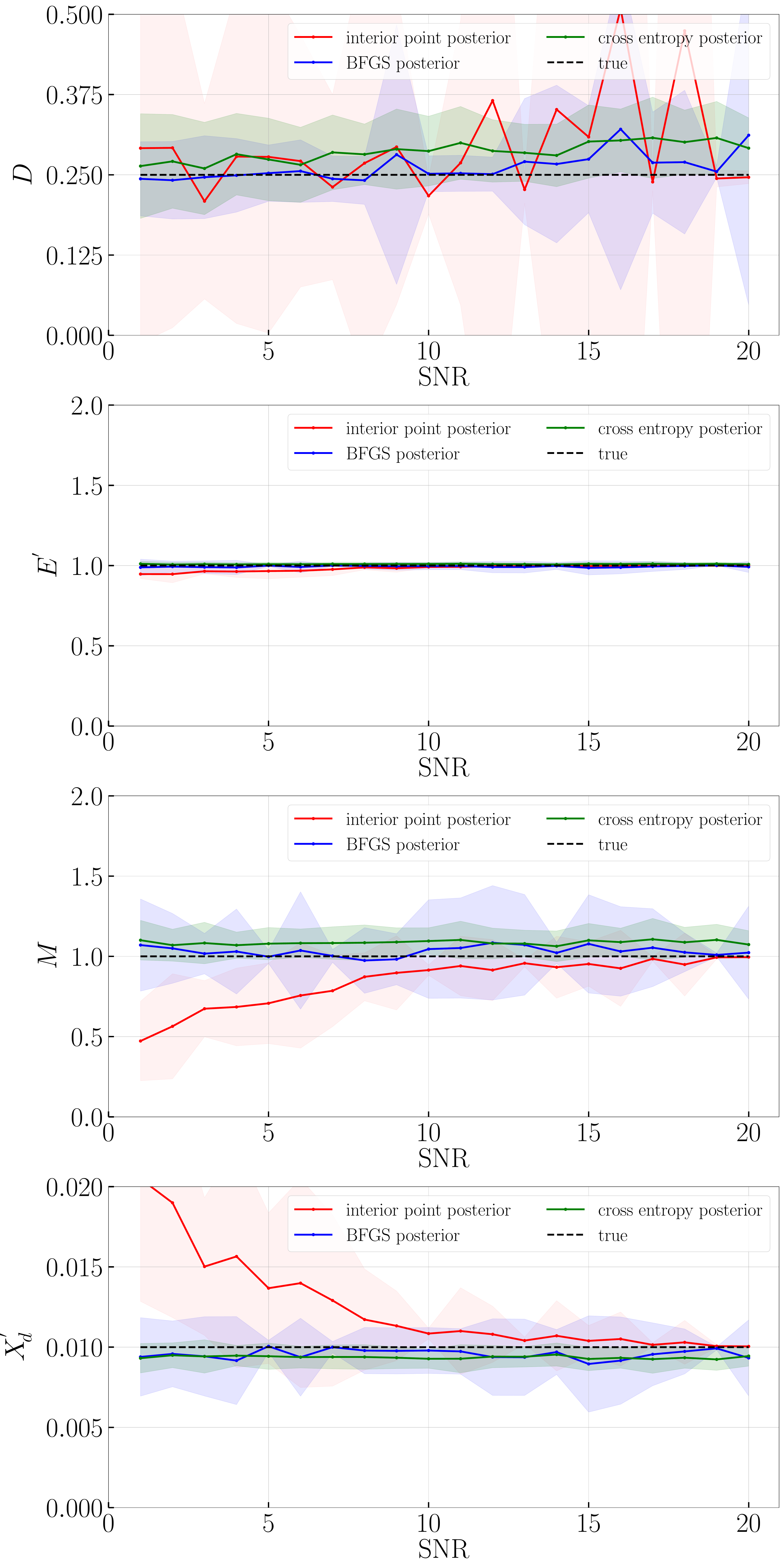}
\caption{Convergence of 4 generator parameters depending on SNR.
Every optimization procedure has been used to obtain 50 posterior values of generator parameters.
Any curve is a mean of these 50 numbers.
Filled area around a curve denotes square root of variance's of unbiased estimation.
}
\label{all_gen_params_convergence}
\end{figure}

A particular advantage of this method as compared to methods requiring computations of the gradient or the Hessian of the objective function \cite{bayesian_framework,gorbunov2019estimation,mackay2019inference} is illustrated in Fig.~\ref{all_gen_params_convergence}. In this experiment, we obtain posterior system parameters $\Theta_{\text{post}}$ starting from their prior values $\Theta_{\text{prior}}$ using different optimization methods in the whole range of the SNRs, from SNR = 1 to SNR  = 20. We compared (1) a modified version of the interior point method proposed for this problem in~\cite{bayesian_framework}; (2) a state-of-the-art first-order optimization method BFGS \cite{liu1989limited,najafabadi2017large}, which often has a superior computational performance in practice; and (3) the cross-entropy method, provided in~Alg.~\ref{ce_algo}. 
For each of the algorithms we computed the average value of the parameters found by the method. Filled area around a curve denotes a standard deviation of unbiased estimation after 50 runs. Further details on the experiments are provided in Section~\ref{sec:emp}.

\begin{algorithm}[!t]
\SetAlgoLined
    \KwIn{objective function ${F}(\Theta)$ to be minimized, \par \qquad prior for the system parameters $\Theta_0$,} 
    \par \qquad smoothing parameter $\alpha$, $\alpha \in (0,1)$
    \par \qquad number of samples generated on each iteration $N$; 
    \par \qquad number of samples used for parameters update $N^e$\\ 
    \KwOut{optimized parameters value $\Theta_t$}
    \textbf{Initialize:} iteration counter $t= 0$; \par 
    \qquad vector ${\sigma}_0$ of expected deviations of the prior \\ 
    \qquad and the optimal values of the parameters, $\Theta_0 - \Theta_*$. \\
    \textbf{while} $\max_j{(\sigma_{tj})}<\varepsilon$: \\
\nl    \hspace{14pt} Generate $N$ random samples\\ \qquad \qquad ${X}_1, \ldots, {X}_N \sim \mathcal{N}({\Theta}_{t}, \sigma_{t}^{2})$\\ 
\nl    \hspace{14pt} Update $t := t+1$;\\
\nl    \hspace{14pt} Set $\mathcal{I}$ be the indices of $N^{\text{e}}$ samples having  the
    \par \qquad\qquad smallest values of~$F(X_i)$ \\
\nl    \hspace{14pt} Update: For all $j=1, \ldots, n$: 
    \begin{equation}
        \Theta_{tj} := \sum_{i \in \mathcal{I}} X_{ij} / N^{e},
    \end{equation}
    \begin{equation}
        \sigma_{tj} := \sum_{i \in \mathcal{I}} (X_{ij} - \Theta_{tj})^2 / N^{e}.
    \end{equation}
    \\ 
\nl    \hspace{14pt} Smooth the estimations: 
    \begin{equation}
    \begin{split}
        {\Theta}_t := \alpha {\Theta}_t + (1-\alpha) {\Theta}_{t-1}, \\
        {\sigma}_t := \alpha {\sigma}_t + (1-\alpha) {\sigma}_{t-1}.
    \end{split}
    \end{equation}
    \\
\caption{The cross-entropy optimization method.}
\label{ce_algo}
\end{algorithm}

The cross-entropy method practically provides much more reliable estimates for the power system component identification problem as compared to first- and second-order optimization techniques. This is partially due to the numerical instability of the gradient and Hessian computations for sufficiently high SNRs, whereas the objective function value computation required by the method is much more reliable and time-efficient. Note that zero-order algorithms, such as the cross-entropy method, are applicable for sufficiently smaller dimensional optimization problems practice (up to 20 parameters). For problems in a higher dimension, the first-order methods can be preferable.


\begin{figure}[!t]
\centering
\includegraphics[width=0.49\textwidth]{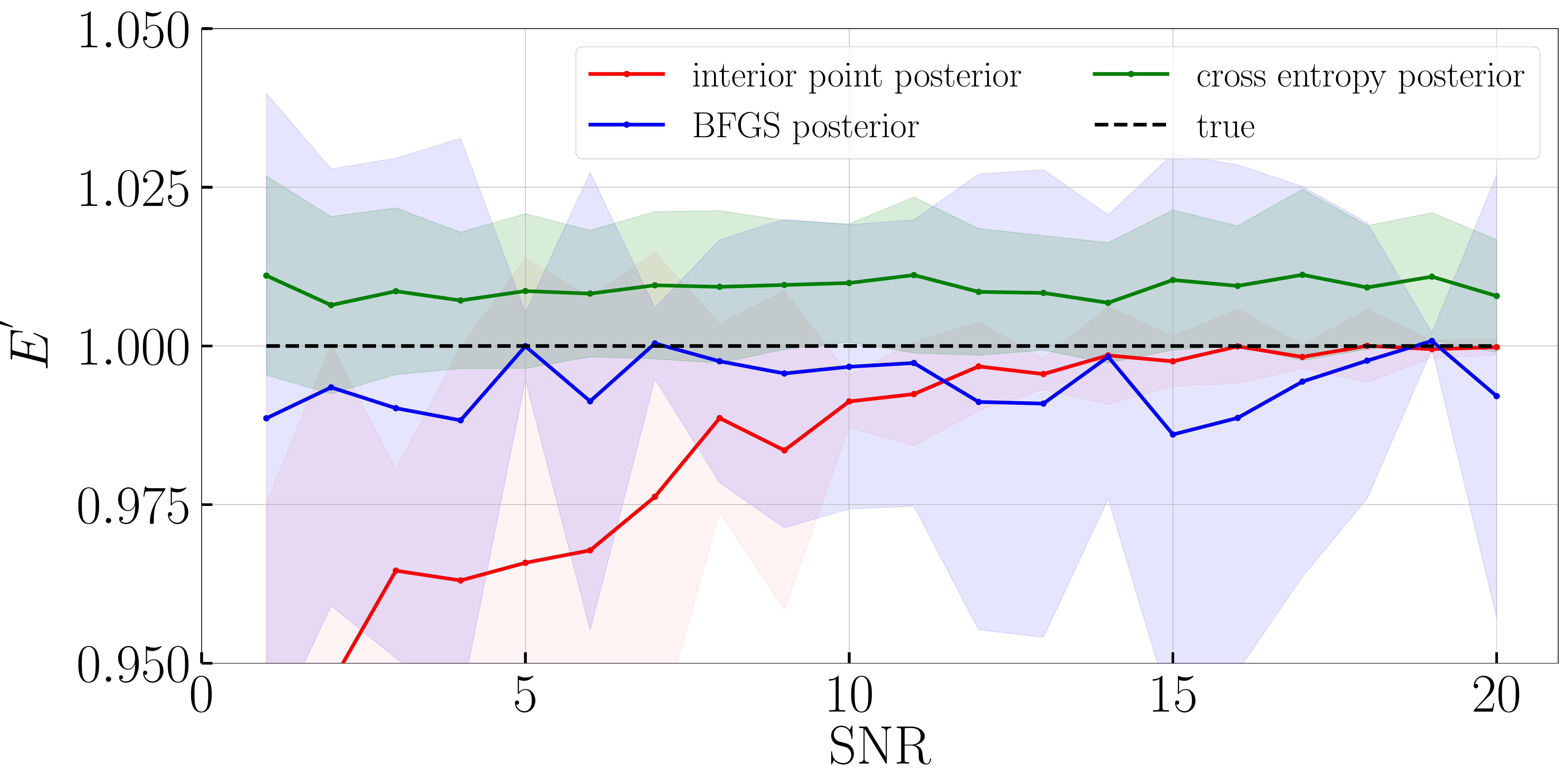}
\caption{Convergence of the $E^{'}$ generator parameter depending on SNR in increased scale.
All plots in Fig.~\ref{all_gen_params_convergence} are shown in the same scale (intervals from 0 to doubled true values).
For this reason, to study convergence of the second parameter, a separate plot should be prepared in increased scale.
}
\label{gen_params_convergence_E}
\end{figure}





\section{Test Results}\label{sec:emp}

\begin{figure}[!ht]
\centering
\includegraphics[width=0.49\textwidth]{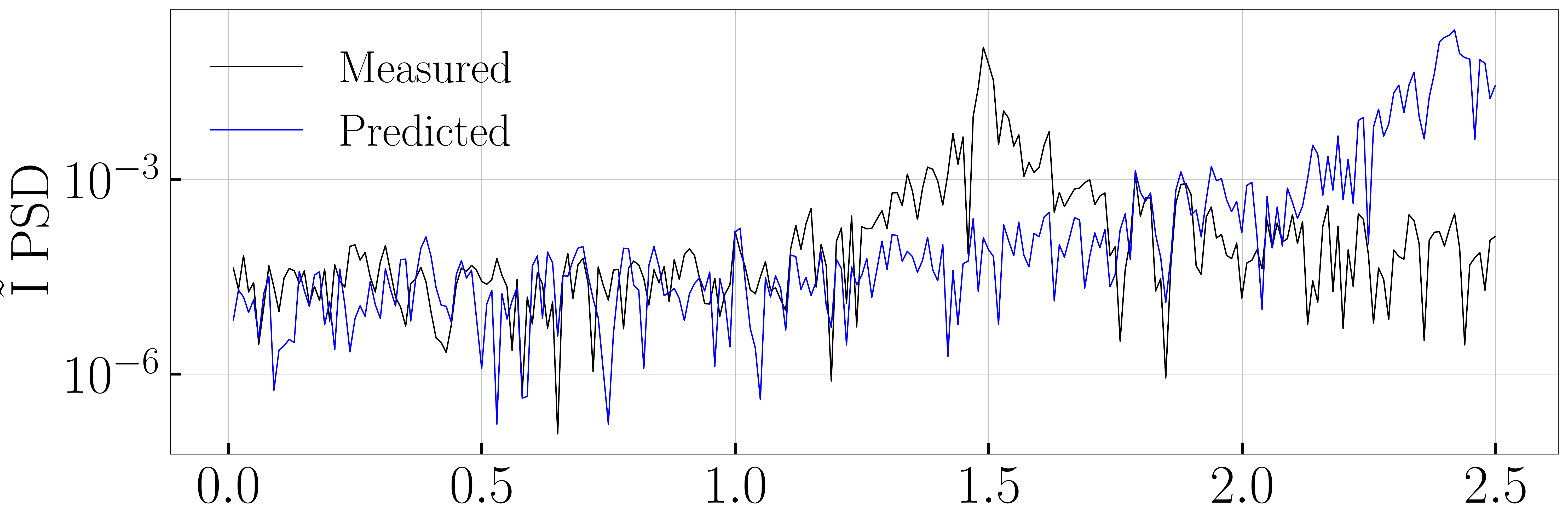}\\
\vspace{4mm}
\includegraphics[width=0.49\textwidth]{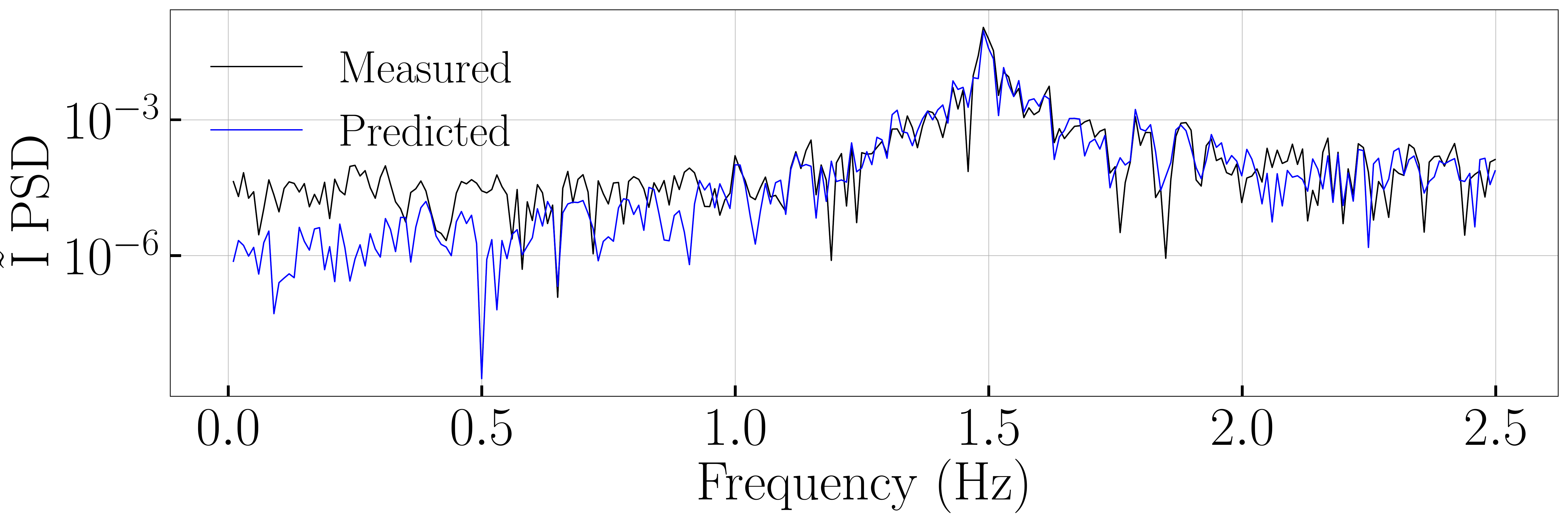}
\caption{Power spectral density of the predicted $\Y(\Theta, \Omega) \tilde{\mathbf{u}}(\Omega)$ and measured $\tilde{\mathbf{y}}(\Omega)$ current magnitude before and after parameter evaluation procedure by maximizing \eqref{Bayes_rule}. The signal-to-noise ratio equals 10.}
\label{Im_PSD}
\end{figure}

Here we present the validation of our proposed framework over a power generator model described in \cite{gen_impedance}. The model contains four generator parameters to be estimated:
damping factor $D$, field voltage magnitude $E^{'}$, inertia $M$, and transient reactance $X^{'}_{d}$.
Thus, the vector of parameters is $ \Theta = [D; E^{'}; M; X^{'}_{d} ]^{\! \top} $.
The true values for these parameters are assumed to have the following values:  $\Theta_{\text{true}}=[0.25; \, 1; \, 1; \, 0.01]^{\! \top}$. Perturbations of the terminal voltage of the generator are assumed to be caused by random load variations of the whole system. This can be represented as white noise variations of the generator terminal voltage. As an illustration, Fig. \ref{Im_PSD} represents the discrepancy between the measured and predicted current power spectral density before and after the parameter inference. The model accuracy is greatly improved by the procedure. 

In engineering practice, the parameters are often not known with high accuracy; therefore, for our tests, we use $50$ scenarios with different priors. We draw them randomly from the range $\pm 50\%$ of the true value for every parameter. We also assume the prior variance of every parameter corresponds to the RMS of $50\%$ of its true value. After that, every scenario is processed at different levels of SNR. The results of identification damping factor $D$, field voltage magnitude $E^{'}$, inertia $M$ and transient reactance $X^{'}_{d}$ --- averaged over all scenarios --- are presented in Fig. \ref{all_gen_params_convergence} for different levels of SNR. 

First, we conclude our method of cross-entropy performs better than the other optimization methods over all SNR values, and especially so for smaller SNR. It outperforms both interior point and BFGS methods both in terms of the bias and the variance of the final predictions. Next, we see that the generator internal EMF can be identified with the most accuracy over the wide range of SNR using any of the methods. This is expected because the components of the generator effective admittance matrix have the most dependence on the generator EMF. For inertia and reactance, the cross-entropy method works best, especially at lower values of SNR. The effective damping is the most difficult parameter to estimate under any SNR, however, our cross-entropy method performs best for this case as well.    

\section{Conclusion and future work}\label{sec:conlusion}

In this paper, we presented a Bayesian framework for identification of dynamic system parameters from measurements of ambient fluctuations of its inputs and outputs. Due to modest signal-to-noise ratios (SNRs) for ambient fluctuation measurements, the maximum likelihood estimation can lead to a rather ill-posed optimization problem. We introduced a special cross-entropy method that does not require expensive gradient or Hessian computations. The method was verified on a model of a synchronous generator under random fluctuations of terminal voltage. We illustrated that our method is superior to first- and second-order methods (gradient-based and interior point methods, respectively) over a wide range of SNR values. In particular, using first- and second-order optimization methods for ill-posed problems often leads to numerical instability and trapping of the method in a spurious local minimum, whereas the cross-entropy method is reliable, even for small values of SNR. 

\bibliographystyle{IEEEtran}
\bibliography{biblio}

\end{document}